\documentclass[a4paper,12pt]{article}

\begin{document}

\author{C. Barrab\`es\thanks{E-mail : barrabes@celfi.phys.univ-tours.fr}
 \ ,\\     
\small Laboratoire de Math\'ematiques et Physique Th\'eorique\\
\small  CNRS/UPRES-A, Universit\'e F. Rabelais, 37200 TOURS, 
France\\ \\G. F. Bressange\thanks{E-mail : georges.bressange@ucd.ie} and 
P.A. Hogan\thanks{E-mail : phogan@ollamh.ucd.ie}\\
\small Mathematical Physics Department\\
\small  University College Dublin, Belfield, Dublin 4, Ireland}

\title{Colliding Plane Impulsive Gravitational Waves}
\date{}
\maketitle

\begin{abstract}
When two non--interacting plane impulsive gravitational 
waves undergo a head--on collision, the vacuum interaction 
region between the waves after the collision contains backscattered 
gravitational radiation from both waves. The two systems of 
backscattered waves have each got a family of rays (null 
geodesics) associated with them. We demonstrate that if it 
is assumed that a parameter exists along each of these 
families of rays such that the modulus of the complex 
shear of each is equal then Einstein's vacuum field 
equations, with the appropriate boundary conditions, 
can be integrated systematically to reveal the 
well--known solutions in the interaction region. 
In so doing the mystery behind the origin of such 
solutions is removed. With the use of the field equations 
it is suggested that the assumption leading to their 
integration may be interpreted physically as implying that 
the energy densities of the two backscattered radiation 
fields are equal. With the use of different boundary 
conditions this approach can lead to new collision 
solutions.
\end{abstract}
\thispagestyle{empty}
\newpage

\section{Introduction}\indent
This paper is a study of the space--time describing the 
vacuum gravitational field left behind after the head--on 
collision of two plane impulsive gravitational waves. The known 
exact solutions are the classical solution of Khan and Penrose 
\cite{KP} and its generalisation by Nutku and Halil \cite{HN}. 
No details of the derivation were given by Khan and Penrose. The 
Nutku and Halil solution was obtained using a harmonic mapping 
technique. The latter solution was subsequently rederived by 
Chandrasekhar and Ferrari \cite{CF} (to quote from \cite{CF}: "This paper is addressed, principally, to a more standard derivation of the Nutku-Halil solution than the one sketched by the authors.") who developed an Ernst--type 
formulation for vacuum space--times admitting two space--like 
Killing vectors. They demonstrated that ``in some sense, the 
Nutku--Halil solution occupies the 
same place in space--times with two space--like Killing vectors as the 
Kerr solution does in space--times with one time--like and one space--like 
Killing vector". Thus the origin of the solution is still quite 
mysterious. Of course the task of solving 
Einstein's vacuum field equations with 
appropriate boundary conditions for the space--time 
after the waves collide is mathematically very complex. 
On the other hand the physical picture, at least up to the 
appearance of a curvature singularity, is quite simple: two 
non--interacting plane impulsive waves undergo a head--on collision and 
the interaction region afterwards contains backscattered radiation 
from both waves, neither of which remain plane. Our aim in this 
paper is to introduce a simple assumption based on this picture 
which provides a key to the origin of the exact solution describing the 
collision of two completely arbitrary plane impulsive gravitational 
waves. 

The backscattered radiation in the interaction region of 
space--time between the histories of the waves after the 
collision determines two intersecting congruences of null 
geodesics. These are the `rays' associated with the two 
systems of backscattered radiation. Both congruences have 
expansion and shear. The ratio of the expansions of each 
congruence is immediately 
determined from Einstein's vacuum field equations and the 
boundary conditions. The shear of each congruence (the modulus of 
a complex variable in each case) depends in general in 
a simple way on the choice of parameter along the null geodesics 
(in the sense that a change of parameter induces a rescaling of 
the shear). {\it We assume that a parameter exists along 
each of the two families of null geodesics such that the 
shear (i.e. the modulus of the `complex shear') of each 
congruence is equal}. This is the only assumption made apart 
from the usual assumption of analyticity of the solution 
of Einstein's field equations \cite{HMAC}. We show that it 
implies, together with the field equations, an equation that 
could be interpreted physically as saying that the energy 
density of the backscattered radiation from each wave 
after collision is the same. In addition we demonstrate how 
this assumption leads to the complete integration of the vacuum 
field equations.

The outline of the paper is as follows: In section 2 the 
collision problem is formulated as a boundary--value problem. 
At this stage, to make the paper as self--contained as possible, 
reference is made to an Appendix A giving a brief summary of 
the construction of a plane impulsive gravitational wave in 
the manner of Penrose \cite{P}. The backscattered radiation 
fields, existing after the collision, are introduced in section 3. The 
assumption central to this study is also introduced in this 
section and a physical implication is explored. Although the 
key assumption is simple the full integration 
of Einstein's vacuum field equations emerging from it is still 
quite complicated and this is described in section 4. The 
complications arise because the incoming waves are not in general linearly 
polarised. For readers who do not want to work through section 4 
the considerably simpler case of linearly polarised incoming 
waves is treated in Appendix B. It is shown there that our 
basic assumption leads to the Khan and Penrose \cite{KP} 
solution. Finally in section 5 the results of section 4 are summarised and 
contact is made with the known exact solutions \cite {KP}, 
\cite {HN}.

With the use of different boundary conditions to those employed 
here the approach of this paper can lead to new collision 
solutions of the field equations. The present authors have 
published a new collision solution of the Einstein--Maxwell 
field equations in \cite {LMP} derived using the ideas 
described in the present paper.

\setcounter{equation}{0}
\section{The Boundary--Value Problem}\indent
The line--element of the space--time describing the vacuum 
gravitational field of a single plane impulsive gravitational 
wave having the maximum two degrees of freedom of 
polarisation may be written in the form (see Appendix A)
\begin{equation}\label{2.1}
ds^2=-2\,\left |d\zeta +v_+(a-ib)\,
d\bar\zeta\,\right |^2+2\,du\,dv\ ,
\end{equation}
where $a, b$ are real constants. Here and throughout this 
paper a bar will denote complex conjugation. The history of the wave is 
the null hyperplane $v=0$ and $v_+=v\,\theta (v)$ where $\theta (v)$ 
is the Heaviside step function. $u$ is a second null coordinate. 
The space--time to the future ($v>0$) of the history of the wave 
is Minkowskian and so is the space--time to the past ($v<0$) 
of $v=0$. Writing $\sqrt{2}\,\zeta =x+iy$ we see that (\ref{2.1}) 
is in the Rosen--Szekeres form
\begin{equation}\label{2.2}
ds^2=-{\rm e}^{-U}\left ({\rm e}^V\cosh W\,dx^2-2\sinh W\,dx\,dy
+{\rm e}^{-V}\cosh W\,dy^2\right )+2\,{\rm e}^{-M}du\,dv\ ,
\end{equation}
with
\begin{eqnarray}\label{2.3}
{\rm e}^{-U}&=& 1-(a^2+b^2)v_+^2\ ,\\\label{2.4}
{\rm e}^{-V}&=& \left [\frac{1+(a^2+b^2)v_+^2-2av_+}{1+(a^2+b^2)v_+^2+2av_+}\right ]
^{\frac{1}{2}}\ ,\\\label{2.5}
\sinh W&=& \frac{2bv_+}{1-(a^2+b^2)v_+^2}\ 
,\\\label{2.6}
M&=& 0.
\end{eqnarray}
We consider the head--on collision of this wave with a wave of 
similar type. This latter wave is described by a space--time with 
line--element (\ref{2.2}) but with  
\begin{eqnarray}\label{2.7}
{\rm e}^{-U}&=& 1-(\alpha ^2+\beta ^2)u_+^2\ ,\\\label{2.8}
{\rm e}^{-V}&=& \left [\frac{1+(\alpha ^2+\beta ^2)u_+^2-2\alpha 
u_+}{1+(\alpha ^2+\beta ^2)u_+^2+2\alpha u_+}\right ]
^{\frac{1}{2}}\ ,\\\label{2.9}
\sinh W&=& \frac{2\beta u_+}{1-(\alpha ^2+\beta ^2)u_+^2}\ 
,\\\label{2.10}
M&=& 0,
\end{eqnarray}
with $\alpha , \beta$ real constants and $u_+=u\,\theta (u)$. The 
history of the wave front is the null hyperplane $u=0$ in this 
case. For the collision we consider the space--time to have 
line--element (\ref{2.2}) with $U, V, W, M$ given by (\ref{2.3})--
(\ref{2.6}) in the region $u<0, v>0$ and given by (\ref{2.7})--
(\ref{2.10}) in the region $v<0, u>0$. The region $u<0, v<0$ 
has line--element (\ref{2.2}) with $U=V=W=M=0$ (which agrees with 
(\ref{2.3})--(\ref{2.10}) when both $v<0$ and $u<0$). The line--element 
in the region $u>0, v>0$ (after the collision) has the form 
(\ref{2.2}) with $U, V, W, M$ functions of $(u, v)$ satisfying the 
O'Brien--Synge \cite{OBS} junction conditions: If $u=0, v>0$ then 
$U, V, W, M$ are given by (\ref{2.3})--(\ref{2.6}) with $v_+=v$ 
and if $v=0, u>0$ then $U, V, W, M$ are given by (\ref{2.7})--(\ref{2.10}) 
with $u_+=u$. Einstein's vacuum field equations have to be 
solved for $U, V, W, M$ in the interaction region ($u>0, v>0$) 
after the collision subject to these boundary (junction) 
conditions. These equations are \cite{JG} (with 
subscripts denoting partial derivatives):
\begin{eqnarray}\label{2.11}
U_{uv}&=& U_u\,U_v\ ,\\\label{2.12}
2V_{uv}&=& U_u\,V_v+U_v\,V_u-2\left (V_u\,W_v+V_v\,W_u\right )
\tanh W\ ,\\\label{2.13}
2W_{uv}&=& U_u\,W_v+U_v\,W_u+2V_u\,V_v\,\sinh W\,\cosh W\ ,\\
\label{2.14}
2U_u\,M_u&=& -2U_{uu}+U_u^2+W_u^2+V_u^2\cosh ^2W\ ,\\\label{2.15}
2U_v\,M_v&=& -2U_{vv}+U_v^2+W_v^2+V_v^2\cosh ^2W\ ,\\\label{2.16}
2M_{uv}&=& -U_{uv}+W_u\,W_v+V_u\,V_v\,\cosh ^2W\ .
\end{eqnarray}
The first of these equations can immediately be solved \cite{EX} in conjuction 
with the boundary conditions to be satisfied by $U$ on $u=0$ 
and on $v=0$ to yield, in $u>0, v>0$,
\begin{equation}\label{2.17}
{\rm e}^{-U}=1-(a^2+b^2)\,v^2-(\alpha ^2+\beta ^2)\,u^2\ .\end{equation}
The problem is to solve (\ref{2.12})--(\ref{2.13}) for $V, W$ 
subject to the boundary conditions and then to solve (\ref{2.14}) 
and (\ref{2.15}) for $M$. Equation (\ref{2.16}) is the integrability 
condition for (\ref{2.14}) and (\ref{2.15}).

\setcounter{equation}{0}
\section{The Backscattered Radiation Fields}\indent 
We shall for the moment focus attention on the two field 
equations (\ref{2.12}) and (\ref{2.13}). All of our 
considerations from now on will apply to the interaction 
region of space--time $u>0, v>0$ after the collision. Introducing 
the complex variables
\begin{equation}\label{3.1}
A=-V_u\cosh W+iW_u\ ,\qquad B=-V_v\cosh W+iW_v\ ,\end{equation}
we can rewrite the two real equations (\ref{2.12}) and (\ref{2.13}) as the 
single complex equation
\begin{equation}\label{3.2}
2A_v=U_u\,B+U_v\,A-2iA\,V_v\sinh W\ ,\end{equation}
or equivalently as the single complex equation
\begin{equation}\label{3.3}
2B_u=U_u\,B+U_v\,A-2iB\,V_u\sinh W\ .\end{equation}
Given the form of the line--element (\ref{2.2}) it is 
convenient to introduce a null tetrad $\{m, \bar m, l, n\}$ 
in the region $u>0, v>0$ defined by
\begin{equation}\label{3.4}
m=\frac{{\rm e}^{U/2}}{\sqrt{2}}\left[{\rm e}^{-V/2}\left (
\cosh\frac{W}{2}-i\sinh\frac{W}{2}\right )\,\frac{\partial}
{\partial x}+{\rm e}^{V/2}\left (\sinh\frac{W}{2}-i\cosh\frac{W}
{2}\right )\,\frac{\partial}{\partial y}\right ]\ ,
\end{equation}
\begin{equation}\label{3.5}
l= {\rm e}^{M/2}\frac{\partial}{\partial v}\ ,
\end{equation}
\begin{equation}\label{3.6}
n= {\rm e}^{M/2}\frac{\partial}{\partial u}\ ,
\end{equation}
with $\bar m$ the complex conjugate of $m$. The integral curves 
of the vector fields $l$ and $n$ are twist--free, null geodesics. 
The coordinate $v$ is not an affine parameter along the 
integral curves of $l$ and these curves have complex shear 
$\sigma _l$ and real expansion $\rho _l$ (we use the 
standard definitions for these quantities given in $\S$4.5 
of \cite{PR} for example) given by
\begin{equation}\label{3.7}
\sigma _l=\frac{1}{2}\,{\rm e}^{M/2}\,B\ ,\qquad \rho _l=\frac{1}{2}\,
{\rm e}^{M/2}\,U_v\ ,\end{equation}
with $B$ as in (\ref{3.1}). Likewise the coordinate $u$ is not 
an affine parameter along the integral curves of $n$ and these 
curves have complex shear $\sigma _n$ and real expansion $\rho _n$ 
given by 
\begin{equation}\label{3.8}
\sigma _n=\frac{1}{2}\,{\rm e}^{M/2}\,A\ ,\qquad \rho _n=\frac{1}{2}\,
{\rm e}^{M/2}\,U_u\ ,\end{equation}
with $A$ as in (\ref{3.1}). We thus see from (\ref{2.17}), 
(\ref{3.7}) and 
(\ref{3.8}) that the ratio $\rho _l/\rho _n$ is now known in the 
region $u>0, v>0$. In terms of the variables introduced above 
the non--identically vanishing scale--invariant components \cite{JG} 
of the Riemann tensor in Newman--Penrose notation are 
$\Psi _0, \Psi _2, \Psi _4$ given by
\begin{eqnarray}\label{3.9}
2\,\Psi _0&=& B_v+\left (M_v-U_v\right )\,B+iB\,V_v\sinh W\ ,\\
\label{3.10}
2\,\Psi _2&=& M_{uv}-\frac{1}{4}\,\left (A\,
\bar B-\bar A\,B\right )\ ,\\
\label{3.11}
2\,\bar\Psi _4&=& A_u+\left (M_u-U_u\right )\,A+iA\,V_u\sinh W\ .
\end{eqnarray}
When these are non--zero we interpret $\Psi _0$ as describing 
radiation, having propagation direction $n$ in space--time, 
backscattered from the wave with history $u=0, v>0$ and 
we interpret $\Psi _4$ as describing radiation, having 
propagation direction $l$ in space--time, backscattered from 
the wave with history $v=0, u>0$. Thus the integral curves of 
the null vector fields $n$ and $l$ are the `rays' associated 
with the backscattered radiation from the two separating 
waves after collision. 

We now look for an interesting assumption to make regarding the 
rays associated with the backscattered radiation fields. Since 
$\rho _l/\rho _n$ is known from (\ref{3.7}), (\ref{3.8}) and 
(\ref{2.17}) we focus attention on the complex shears 
$\sigma _l$ and $\sigma _n$ in (\ref{3.7}) and (\ref{3.8}). Let 
us write
\begin{equation}\label{3.12}
A=\left |A\right |\,{\rm e}^{i\theta}\ ,\qquad 
B=\left |B\right |\,{\rm e}^{i\phi}\ ,\qquad 
f=\theta -\phi\ ,\end{equation}
with $\theta$ and $\phi$ real. From (\ref{3.2}) and (\ref{3.3}) 
we can obtain the equations
\begin{eqnarray}\label{3.13}
\theta _v&=& -\frac{\left |B\right |}{2\left |A\right |}\,U_u\,\sin f
-V_v\sinh W\ ,\\
\label{3.14}
 \phi _u&=& \frac{\left |A\right |}{2\left |B\right |}\,U_v\,\sin f
-V_u\sinh W\ ,\end{eqnarray}
and
\begin{equation}\label{3.15}
2f_{uv}+i\left (A\,\bar B-\bar A\,B\right )=-\left (U_u\,
\frac{\left |B\right |}{\left |A\right |}\,\sin f\right )_u
-\left (U_v\,
\frac{\left |A\right |}{\left |B\right |}\,\sin f\right )_v\ .
\end{equation}
The arguments $\theta , \phi$ of $A, B$ respectively are tetrad 
dependent. If we transform the tetrad $\{m, \bar m, l, n\}$ by the 
rotation
\begin{equation}\label{3.16}
m\rightarrow \hat m={\rm e}^{i\psi}\,m\ ,\end{equation}
with $\psi$ a real--valued function, then
\begin{equation}\label{3.17}
\theta\rightarrow\hat\theta =\theta -2\,\psi\ ,\qquad 
\phi\rightarrow\hat\phi =\phi -2\,\psi\ ,\qquad f\rightarrow f\ .
\end{equation}
On account of the field equation (\ref{3.15}) {\it we can, 
without loss of generality, arrange to have  $\theta +\phi =
{\rm constant}$}. To see this we deduce from 
(\ref{3.13}) and (\ref{3.14}) that 
\begin{eqnarray}\label{3.18}
\left (\hat\theta +\hat\phi\right )_u&=& \frac{\left |A\right |}{\left |B
\right |}\,U_v\,\sin f+f_u-2\,V_u\sinh W-4\psi _u\ ,\\
\label{3.19}
\left (\hat\theta +\hat\phi\right )_v&=& -\frac{\left |B\right |}
{\left |A\right |}\,U_u\,\sin f-f_v-2\,V_v\sinh W-4\psi _v\ .
\end{eqnarray}
We are free to choose $\psi$ to make the right hand sides of 
(\ref{3.18}) and (\ref{3.19}) vanish because the integrability 
condition for the resulting pair of first order partial 
differential equations for $\psi$ is the field equation (\ref{3.15}). 
Hence it is always possible to choose a tetrad so that $\theta$ 
and $\phi$ in (\ref{3.12}) have the property that $\theta +\phi =
{\rm constant}$. This result suggests that to discover an 
interesting assumption to make 
about the rays associated with the backscattered radiation fields 
we should consider the ratio (because it depends upon $\theta -\phi$ 
and not $\theta +\phi$) 
\begin{equation}\label{3.20}
\frac{A}{B}=\frac{\left |A\right |}{\left |B\right |}\,
{\rm e}^{if}=\frac{\sigma _n}{\sigma _l}\ ,\end{equation}
with the last equality following from (\ref{3.7}) and (\ref{3.8}). 
We note that $f$ satisfies the second order equation (\ref{3.15}) 
to which we shall return later. It is clear from (\ref{3.1}) that 
a change 
of parameters $u\rightarrow\bar u=\bar u(u)$ and 
$v\rightarrow\bar v
=\bar v(v)$ along the integral curves of $n$ and $l$ rescales 
$A$ and $B$ by a function of $u$ and a function of $v$ respectively. 
This change of parameter obviously leaves the form of the line--element 
(\ref{2.2}) invariant. Also from the field equation (\ref{3.2}) we deduce 
that
\begin{equation}\label{3.21}
\left (\left |A\right |^2\right )_v-U_v\,\left |A\right |^2=
\frac{1}{2}\,U_u\,\left (A\,\bar B+\bar A\,B\right )\ ,
\end{equation}
and from the equivalent equation (\ref{3.3}) we find 
\begin{equation}\label{3.22}
\left (\left |B\right |^2\right )_u-U_u\,\left |B\right |^2=
\frac{1}{2}\,U_v\,\left (A\,\bar B+\bar A\,B\right )\ ,
\end{equation}
>From these two equations we obtain
\begin{equation}\label{3.23}
2\,\left [{\rm log}\left (\frac{\left |A\right |^2}{\left |
B\right |^2}\right )\right ]_{uv}=\left (U_u\,
\frac{\left (A\,\bar B+\bar A\,B\right )}{\left |A\right |
^2}\right )_u-\left (U_v\,
\frac{\left (A\,\bar B+\bar A\,B\right )}{\left |B\right |
^2}\right )_v\ ,\end{equation}
which is a partner for the equation (\ref{3.15}) for 
$f$. This suggests that it might be interesting to explore the 
following assumption concerning the rays associated 
with the backscattered radiation: {\it there exist 
parameters $\bar u, \bar v$ along the integral 
curves of $n$ and $l$ respectively such that $\left |
A\right |^2=\left |B\right |^2$ (or equivalently 
$\left |\sigma _n\right |^2=\left |\sigma _l\right |^2$)}. This 
is equivalent to the assumption that there exist functions $C(u), D(v)$ 
such that 
\begin{equation}\label{3.24}
\frac{\left |A\right |^2}{\left |B\right |^2}=
C(u)\,D(v)\ .\end{equation}
When $u=0$ it follows from (\ref{2.4}) and (\ref{2.5}) that 
\begin{equation}\label{3.25}
\left |B\right |^2=\frac{4\,(a^2+b^2)}{\left (1-(a^2+b^2)\,v^2\right )^2}
\ ,\end{equation}
and when $v=0$ it follows from (\ref{2.8}) and (\ref{2.9}) that 
\begin{equation}\label{3.26}
\left |A\right |^2=\frac{4\,(\alpha ^2+\beta ^2)}{\left (1-(\alpha ^2+\beta ^2)\,u^2\right )^2}
\ .\end{equation}
Also when $u=0$ we see from (\ref{2.17}) that the right hand 
side of (\ref{3.21}) vanishes and thus solving (\ref{3.21}) 
for $\left |A\right |^2$ when $u=0$ we obtain 
\begin{equation}\label{3.27}
\left |A\right |^2=\frac{4\,(\alpha ^2+\beta ^2)}{1-(a^2+b^2)\,v^2}\ ,
\end{equation}
with the constant numerator here (the constant of integration) 
chosen so that the two expressions (\ref{3.26}) and (\ref{3.27}) 
for $\left |A\right |^2$ agree when $u=0$ {\it and} $v=0$. 
Similarly when $v=0$ the right hand side of (\ref{3.22}) 
vanishes and we readily obtain, when $v=0$,
\begin{equation}\label{3.28}
\left |B\right |^2=\frac{4\,(a^2+b^2)}{1-(\alpha ^2+\beta ^2)\,u^2}
\ .\end{equation}
Thus (\ref{3.24}) together with the boundary conditions at $u=0$ 
and at $v=0$ results in 
\begin{equation}\label{3.29}
\frac{\left |A\right |^2}{\left |B\right |^2}=
\left (\frac{\alpha ^2+\beta ^2}{a^2+b^2}\right )\,
\left [\frac{1-(a^2+b^2)\,v^2}{1-(\alpha ^2+\beta ^2)\,u^2}
\right ]\ .\end{equation}
Hence there exist parameters $(\bar u, \bar v)$ given by 
\begin{equation}\label{3.30}
\bar u=\sin ^{-1}\left (u\,\sqrt{\alpha ^2+\beta ^2}\right )\ ,
\qquad \bar v=\sin ^{-1}\left (v\,\sqrt{a^2+b^2}\right )\ ,
\end{equation}
such that 
\begin{equation}\label{3.31}
\frac{V_{\bar u}^2\cosh ^2W+W_{\bar u}^2}
{V_{\bar v}^2\cosh ^2W+W_{\bar v}^2}=1\ .\end{equation}
We note that when $\bar u=0$ we have $u=0$ and when $\bar v=0$ 
we have $v=0$. We shall express (\ref{3.31}) 
by saying that, in the coordinates $(\bar u, \bar v)$, 
(\ref{3.29}) reads $\left |A\right |^2=\left |B\right |^2$. 
In the coordinates $(\bar u, \bar v)$ the form of the 
field equations and the expressions for the Riemann tensor 
remain invariant, with the derivatives with respect to 
$(u, v)$ being replaced by derivatives with respect to $(\bar u, \bar v)$ 
and with $M$ replaced by $\bar M$ according to
\begin{equation}\label{3.32}
{\rm e}^{-\bar M}=\frac{\cos\bar u\,\cos\bar v}{\sqrt{\alpha ^2+
\beta ^2}\,\sqrt{a^2+b^2}}\,{\rm e}^{-M}\ .\end{equation}
We note from (\ref{2.17}) and (\ref{3.30}) that in the barred 
coordinates
\begin{equation}\label{3.33}
{\rm e}^{-U}=\cos (\bar u-\bar v)\,\cos (\bar u+\bar v)\ .
\end{equation}
Also in these coordinates (\ref{3.23}) becomes
\begin{equation}\label{3.34}
\left (U_{\bar u}\,\cos f\right )_{\bar u}=\left (U_{\bar v}\,
\cos f \right )_{\bar v}\qquad\Longleftrightarrow
\qquad U_{\bar u}\,f_{\bar u}
=U_{\bar v}\,f_{\bar v}\ ,
\end{equation} 
(the equivalence here following from (\ref{3.33}) since now 
$U_{\bar u\bar u}=U_{\bar v\bar v}$) from which we conclude 
that
\begin{equation}\label{3.35}
f=f(\lambda )\qquad {\rm with}\qquad \lambda =\frac{\cos (\bar u-\bar v)}
{\cos (\bar u+\bar v)}\ .\end{equation}
We see that $\lambda =1$ when $\bar u=0$ and/or when $\bar v=0$. 
Also (\ref{3.21}) and (\ref{3.22}) become 
\begin{eqnarray}\label{3.36}
\left (\left |A\right |^2\right )_{\bar v}-U_{\bar v}\,
\left |A\right |^2&=& U_{\bar u}\,\left |A\right |^2\,\cos f\ ,\\
\label{3.37}
\left (\left |A\right |^2\right )_{\bar u}-U_{\bar u}\,
\left |A\right |^2&=& U_{\bar v}\,\left |A\right |^2\,\cos f\ .
\end{eqnarray}
It thus follows that, again in the barred coordinates,
\begin{equation}\label{3.38}
\left |A\right |^2=\left |B\right |^2={\rm e}^U\,g(\lambda )\ ,
\end{equation}
for some function $g(\lambda )$ satisfying
\begin{equation}\label{3.39}
\lambda\,g'=g\,\cos f\ ,\end{equation}
with the prime here and henceforth denoting differentiation 
with respect to $\lambda$.

In order to intepret physically the implications of 
our assumption that there 
exists $(\bar u, \bar v)$ such that $\left |A\right |^2=\left |
B\right |^2$ we proceed as follows: using (\ref{3.20}) 
and the field equations (\ref{2.14}) and (\ref{2.15}) 
in the expressions (\ref{3.9}) and (\ref{3.11}) 
for $\Psi _0$ and $\Psi _4$ we find that we can 
write
\begin{equation}\label{3.40}
U_v\,\frac{\Psi _0}{B}=\frac{1}{2}\,\left\{-iU_v\,f_v
-U_{vv}+\frac{1}{2}\,\left |B\right |^2+\frac{B}{2\,A}
\,U_u\,U_v+U_v\,\left (
{\rm log}\frac{\left |B\right |}{\left |A\right |}\right 
)_v\right\}\ ,\end{equation}
and 
\begin{equation}\label{3.41} 
U_u\,\frac{\bar\Psi _4}{A}=\frac{1}{2}\,
\left\{iU_u\,f_u
-U_{uu}+\frac{1}{2}\,\left |A\right |^2+\frac{A}{2\,B}
\,U_u\,U_v+U_u\,\left (
{\rm log}\frac{\left |A\right |}{\left |B\right |}\right 
)_u\right\}\ .\end{equation}
When these are expressed in the coordinates $(\bar u, \bar v)$ 
we can put $\left |A\right |^2=\left |B\right |^2$ and as 
above $U_{\bar u\bar u}=U_{\bar v\bar v}$ and $U_{\bar u}\,
f_{\bar u}=U_{\bar v}\,f_{\bar v}$ and thus  
\begin{equation}\label{3.42}
U_{\bar u}\,\frac{\Psi _4}{\bar A}=U_{\bar v}\,
\frac{\Psi _0}{B}\ .\end{equation}
>From (\ref{3.42}) it follows, using the second of (\ref{3.7}) 
and of (\ref{3.8}) that
\begin{equation}\label{3.43}
\rho ^{-2}_l\,\left |\Psi _4\right |^2=
\rho ^{-2}_n\,\left |\Psi _0\right |^2\ .\end{equation}
This equation, which is a consequence of our basic assumption, 
deserves some physical interpretation. $\left |\Psi _0\right |^2$ 
and $\left |\Psi _4\right |^2$ are analogous to the energy densities 
of electromagnetic waves propagating in the $n$ and $l$ directions 
respectively in space--time, in the same sense that the Bel--
Robinson tensor \cite {PR} is analogous to the electromagnetic 
energy--momentum tensor. However with the coordinates carrying 
the dimensions of length these quantities have the dimensions 
of $({\rm length})^{-4}$. The quantities $\rho _l^{-2}\left |
\Psi _4\right |^2$ and $\rho _n^{-2}\left |
\Psi _0\right |^2$ both have dimensions 
$({\rm length})^{-2}$ of energy density and are positive definite 
expressions in terms of the backscattered radiation fields [we 
note that the backscattered radiation has non--vanishing shear 
and expansion and thus does not consist of systems of plane waves]. 
Hence it seems reasonable to suggest the following interpretation 
for the equation (\ref{3.43}): {\it in the coordinate system (the barred system) 
in which $\left |A\right |^2=\left |B\right |^2$ the 
energy density of the backscattered radiation from 
each of the separating waves after the collision is 
the same}. We note that (\ref{3.43}) also holds for our 
solution \cite{LMP} which describes the collision of an 
impulsive gravitational wave with an impulsive gravitational 
wave sharing its wave front with an electromagnetic shock 
wave. This latter solution contains the Khan and Penrose \cite{KP} 
solution and a solution of Griffiths \cite{G} as special 
cases. Also (\ref{3.43}) holds (trivially) for the Bell and 
Szekeres \cite{BS} solution describing the collision of two electromagnetic 
shock waves. These examples demonstrate that (\ref{3.43}) 
holds for a class of collision problems involving gravitational 
impulse waves and/or electromagnetic shock waves.

\setcounter{equation}{0}
\section{Integration of the Field Equations}\indent
We begin by writing (\ref{3.15}) in the barred system 
$(\bar u, \bar v)$. Using (\ref{3.20}) with $\left |A\right |=
\left |B\right |$, (\ref{3.35}) and (\ref{3.38}) we find that 
\begin{equation}\label{4.1}
(1-\lambda ^2)\,f''-2\,\lambda\,f'+\frac{g}{\lambda}\,
\sin f=\frac{(1+\lambda ^2)}{\lambda ^2}\,\sin f
-\frac{(1-\lambda ^2)}{\lambda}\,f'\,\cos f\ ,
\end{equation}
with $g$ given in terms of $f$ by (\ref{3.39}). We can 
simplify (\ref{4.1}) to read
\begin{equation}\label{4.2}
g=-\frac{\lambda}{\sin f}\,\frac{d}{d\lambda}\left 
[(1-\lambda ^2)\,\left (f'+\frac{\sin f}{\lambda}\right )\right ]
\ .\end{equation}
We get a single third order equation for $f$ by 
eliminating $g$ (taken to be non--zero) between (\ref{3.39}) 
and (\ref{4.2}). Since we are working in the barred 
coordinate system (\ref{3.20}) gives
\begin{equation}\label{4.3}
\frac{A}{B}=\frac{-V_{\bar u}\cosh W+iW_{\bar u}}
{-V_{\bar v}\cosh W+iW_{\bar v}}={\rm e}^{if}=
\frac{1-ih}{1+ih}\ ,\end{equation}
where, for convenience, we have introduced $h(\lambda)$ by the 
final equality. After eliminating $g$ from (\ref{3.39}) and 
(\ref{4.2}) it is useful to write the resulting 
equation as a differential equation for $h(\lambda )$.
Then defining 
\begin{equation}\label{4.4}
G=-\frac{2}{1+h^2}\,(1-\lambda ^2)\,\left (h'-\frac{h}{\lambda }\right )
\ ,\end{equation}
the equation for $h(\lambda )$ can be put in the form
\begin{equation}\label{4.5}
\lambda\,G''+G'-\frac{4}{\lambda }\,G=-\frac{G\,Q}{
1-\lambda ^2}\ ,\end{equation}
where
\begin{equation}\label{4.6}
Q=\frac{\lambda}{2h}(1-h^2)\,G'+\frac{1}{h}\,
(1+h^2)\,G\ .
\end{equation}
We remark that if we define
\begin{equation}\label{4.7}
P=\frac{\lambda}{2h}(1+h^2)\,G'+\frac{1}{h}\,
(1-h^2)\,G\ ,\end{equation}
then 
\begin{equation}\label{4.8}
\lambda\,P'=Q\qquad {\rm and}\qquad P^2-Q^2=
\lambda ^2\left (G'\right )^2-4\,G^2\ .\end{equation}
In studying the differential equation (\ref{4.5}) 
for $h$ we found it helpful to write (\ref{4.5}) and the 
second of (\ref{4.8}) in the form
\begin{eqnarray}\label{4.9}
\lambda\,G''+G'-\frac{4}{\lambda }\,G&=& -\frac{
\lambda\,G\,P'}{
1-\lambda ^2}\ ,\\
\label{4.10}
P^2-\lambda ^2\left (P'\right )^2&=
& \lambda ^2\left (G'\right )^2-4\,G^2\ ,\end{eqnarray}
and to work with these equations. Before proceeding further 
however we need to know $h$ and $h'$ when $\bar u=0$ 
and/or when $\bar v=0$, i.e. we require $h(1)$ and 
$h'(1)$. To find $h(1)$ start by writing (\ref{4.3}), 
using (\ref{3.30}), as 
\begin{equation}\label{4.11}
\frac{1-i\,h(\lambda)}{1+i\,h(\lambda )}=
\frac{\sqrt{a^2+b^2}\,\sqrt{1-(\alpha ^2+\beta ^2)\,
u^2}}{\sqrt{\alpha ^2+\beta ^2}\,\sqrt{1-(a^2+b^2)\,
v^2}}\,\left (\frac{-V_u\cosh W+iW_u}{-V_v\cosh W
+iW_v}\right )\ ,\end{equation}
and evaluate this equation when $u=0$ {\it and} 
$v=0$. From the boundary conditions on $V$ and $W$ 
given by (\ref{2.4}), (\ref{2.5}), (\ref{2.8}) 
and (\ref{2.9}) we have that when $u=0$ {\it and} 
$v=0$:
\begin{equation}\label{4.12}
V_u=2\,\alpha\ ,V_v=2\,a\ ,W_u=2\,\beta , W_v=2\,b\ ,\end{equation}
and thus from (\ref{4.11}),
\begin{equation}\label{4.13}
\frac{1-i\,h(1)}{1+i\,h(1)}=\frac{\sqrt{a^2+b^2}}{\sqrt{\alpha ^2+\beta ^2}}
\,\left (\frac{\alpha -i\beta}{a-ib}\right )=
{\rm e}^{i(\hat\alpha -\hat\beta)}\ ,\end{equation}
where
\begin{equation}\label{4.14}
{\rm e}^{i\hat\alpha}=\frac{\alpha -i\beta}{
\sqrt{\alpha ^2+\beta ^2}}\ ,\qquad 
{\rm e}^{i\hat\beta}=\frac{a-ib}{
\sqrt{a^2+b^2}}\ .\end{equation}
It thus follows from (\ref{4.13}) that 
\begin{equation}\label{4.15}
h(1)=-\tan\left (\frac{\hat\alpha -
\hat\beta}{2}\right )=k\ ({\rm say})\ .\end{equation}
Next to find $h'(1)$ we begin with (\ref{4.11}) and 
by two differentiations obtain from it
\begin{equation}\label{4.16}
-\frac{4i\,(\alpha ^2+\beta ^2)\,h'(1)}
{\left (1+i\,h(1)\right )^2}=\left [\frac{
\partial ^2}{\partial u\partial v}\left (\frac{
-V_u\cosh W+iW_u}{-V_v\cosh W+iW_v}\right )
\right ]_{(u=0, v=0)}\ .
\end{equation}
To evaluate the right hand side here we first note from 
(\ref{2.17}) that when $u=0$ {\it and} $v=0$, 
$U_u=0$ and $U_v=0$. Also from the boundary 
conditions on $W$ we have $W=0$ when $u=0$ {\it and} 
$v=0$. Now evaluating the field equations (\ref{2.12}) and (\ref{2.13}) 
when $u=0$ {\it and} $v=0$ we easily see that in this 
case
\begin{equation}\label{4.17}
V_{uv}=0\ ,\qquad W_{uv}=0\ .\end{equation}
>From the boundary conditions satisfied by $V$ and $W$ we 
have, when $u=0$ {\it and} $v=0$:
\begin{equation}\label{4.18}
V_{vv}=V_{uu}=W_{vv}=W_{uu}=0\ .\end{equation}
Next differentiating (\ref{2.12}) and (\ref{2.13}) 
with respect to $u$ we find that when $u=0$ {\it and} 
$v=0$:
\begin{equation}\label{4.19}
V_{uvu}=2\,\alpha ^2a-6\,\beta ^2a-8\,b\,\alpha\,\beta\ 
,\qquad W_{uvu}=2\,b\,(\alpha ^2+\beta ^2)+8\,a\,\alpha\,
\beta \ .\end{equation}
Finally differentiating (\ref{2.12}) and (\ref{2.13}) 
with respect to $v$ we find that when $u=0$ {\it and} 
$v=0$:
\begin{equation}\label{4.20}
V_{uvv}=2\,\alpha\,a^2-6\,\alpha\,b^2-8\,a\,b\,\beta\ 
,\qquad W_{uvv}=2\,\beta\,(a^2+b^2)+8\,\alpha\,a\,b\ .
\end{equation}
Now substituting all of these results into the right 
hand side of (\ref{4.16}) we obtain
\begin{equation}\label{4.21}
h'(1)=k\ ,\end{equation}
with $k$ given by (\ref{4.15}).
Using (\ref{4.15}) and (\ref{4.21}) in (\ref{4.4}) 
we see that
\begin{equation}\label{4.22}
G(1)=0=G'(1)\ .\end{equation}
We can now set about solving (\ref{4.9}) and 
(\ref{4.10}) 
for $G$ and then obtain $h(\lambda )$ from 
(\ref{4.4}).

Differentiating (\ref{4.10}) with respect to 
$\lambda $ and using (\ref{4.9}) we find that 
either (a) $P'=0$ or (b) if $P'\neq 0$ then 
\begin{equation}\label{4.23}
\lambda\,P''+P'-\frac{1}{\lambda}\,P=\frac{\lambda\,G\,G'}
{1-\lambda ^2}\ .\end{equation}
We can quickly dispose of case (a). If $P={\rm constant}\neq 0$ 
then (\ref{4.10}) can be integrated to yield
\begin{equation}\label{4.24}
G=\frac{P}{4}\,\left (c_0^2\lambda ^{\pm 2}-\frac{1}
{c_0^2\lambda ^{\pm 2}}\right )\ ,\end{equation}
where $c_0$ is a constant of integration. It is easy to see 
that this constant cannot be chosen to satisfy both boundary 
conditions (4.22). Also if $P=0$ then (\ref{4.10}) integrates 
to 
\begin{equation}\label{4.25}
G=c_1\,\lambda ^{\pm 2}\ ,\end{equation}
with $c_1$ a constant of integration. Clearly we must have 
$c_1=0$ to satisfy (\ref{4.22}). Thus the only acceptable 
solution in case (a) is $G=0$. Turning now to case (b) with 
(\ref{4.23}) holding we find that we can integrate this 
equation once (using (\ref{4.22})) to read \cite{INT}
\begin{equation}\label{4.26}
\lambda\,P'+\left (\frac{1+\lambda ^2}{1-\lambda ^2}\right )\,
P=\frac{\lambda\,G^2}{2\,(1-\lambda ^2)}\ .
\end{equation}
>From the first of (\ref{4.8}) this can be written
\begin{equation}\label{4.27}
\left (P+Q\right )+\lambda ^2\left (P-Q\right )=
\frac{\lambda}{2}\,G^2\ ,\end{equation}
and from (\ref{4.7}) and (\ref{4.8}) this reads
\begin{equation}\label{4.28}
\lambda\,G'=\frac{\lambda\,h}{2\,(1+\lambda ^2h^2)}\,G^2
-2\,\frac{(1-\lambda ^2h^2)}{(1+\lambda ^2h^2)}\,G\ .
\end{equation}
A glance at (\ref{4.3}) and (\ref{4.4}) shows how $h$ and thence $G$ 
are constructed from the functions $V$ and $W$ appearing in 
the line--element (\ref{2.2}) for $u>0, v>0$. On the boundaries 
of this region $\lambda =1$ and within this region $\lambda >1$ 
with $\lambda$ becoming infinite when the right hand side 
of (\ref{2.17}) vanishes. The $\lambda ={\rm constant}>1$ curves 
densely fill the interior of the region $\cal S$(say) with boundaries ($b_1$) 
$u=0, v>0$, ($b_2$) $v=0, u>0$ and ($b_3$) the right hand 
side of (\ref{2.17}) vanishing with $u>0, v>0$. Within $\cal S$ 
there is one curve $\lambda ={\rm constant}>1$ passing 
through each point. When the field equations are completely 
integrated for $(u, v)\epsilon \cal S$ the boundary ($b_3$) turns out 
to be a curvature singularity. For $G$ analytic in $\cal S$ 
we conclude from (\ref{4.22}) and (\ref{4.28}) that 
$G\equiv 0$ in $\cal S$. It thus follows from (\ref{4.4}) 
with the boundary condition (\ref{4.15}) that
\begin{equation}\label{4.29}
h(\lambda )=k\,\lambda\ ,\end{equation}
for $(u, v)\epsilon\cal S$.

We are now at the following stage in the integration of the 
field equations: the function $U$ in the line--element (\ref{2.2}) 
is given by (\ref{2.17}) in coordinates $(u, v)$ or by (\ref{3.33}) 
in coordinates $(\bar u, \bar v)$. Also on account of (\ref{4.3}) 
and (\ref{4.29}) the functions $V$ and $W$ in (\ref{2.2}) satisfy 
the differential equation
\begin{equation}\label{4.30}
\frac{A}{B}=\frac{-V_{\bar u}\cosh W+iW_{\bar u}}
{-V_{\bar v}\cosh W+iW_{\bar v}}=\frac{1-i\,k\,\lambda}
{1+i\,k\,\lambda}\ ,\end{equation}
with $\lambda $ given by (\ref{3.35}) and $k$ by (\ref{4.15}). 
We shall now solve this complex equation for $V$ and $W$ in 
terms of the barred coordinates. First we need to note the 
boundary values of $V$ and $W$ in terms of the barred 
coordinates. By (\ref{2.4}) and (\ref{3.30}) we have when 
$\bar u=0$,
\begin{equation}\label{4.31}
{\rm e}^{-V}=\left [\frac{\left (\sqrt{a^2+b^2}-a\,\sin\bar v\right )^2+b^2
\sin ^2\bar v}{\left (\sqrt{a^2+b^2}+a\,\sin\bar v\right )^2+b^2
\sin ^2\bar v}\right ]^{\frac{1}{2}}=
\frac{\left |1-{\rm e}^{i\hat\beta}\sin\bar v\right |}
{\left |1+{\rm e}^{i\hat\beta}\sin\bar v\right |}\ ,
\end{equation}
with the second equality following from (\ref{4.14}). By 
(\ref{2.5}) and (\ref{3.30}) we have when $\bar u=0$
\begin{equation}\label{4.32}
\sinh W=\frac{2\,b\,\sin\bar v}{\sqrt{a^2+b^2}\,\cos ^2
\bar v}=-i\,\frac{\left ({\rm e}^{-i\hat\beta}\sin\bar v-
{\rm e}^{i\hat\beta}\sin\bar v\right )}{1-\left |{\rm e}
^{-i\hat\beta}\sin\bar v\right |^2}\ .
\end{equation} 
The corresponding boundary values on $\bar v=0$ are 
obtained by replacing $\bar v$ by $\bar u$ and 
$\hat\beta$ by $\hat\alpha$ in the final expressions 
in (\ref{4.31}) and (\ref{4.32}). It is convenient 
to use a complex function $E$ (the Ernst function) 
in place of the two real functions $V$ and $W$ defined 
(in a way that is suggested by the final expressions 
in (\ref{4.31}) and (\ref{4.32})) by
\begin{equation}\label{4.33}
{\rm e}^{-V}=\left [\frac{\left (1-E\right )\,\left (
1-\bar E\right )}{\left (1+E\right )\,\left (
1+\bar E\right )}\right ]^{\frac{1}{2}}\ ,\qquad 
\sinh W=-i\,\frac{\left (E-\bar E\right )}{1-\left |E\right |
^2}\ ,\end{equation}
or equivalently by
\begin{equation}\label{4.34}
E=\frac{\sinh V\,\cosh W+i\sinh W}{1+\cosh V\,\cosh W}\ .
\end{equation}
Now (\ref{4.31}) and (\ref{4.32}) can be written neatly as:
\begin{equation}\label{4.35}
{\rm when}\ \ \bar u=0\ ,\qquad E={\rm e}^{-i\hat\beta}\sin\bar v\ ,
\end{equation}
and correspondingly 
\begin{equation}\label{4.36}
{\rm when}\ \ \bar v=0\ ,\qquad E={\rm e}^{-i\hat\alpha}\sin\bar u
\ .\end{equation}
In terms of $E$, the complex functions $A, B$ can be written
\begin{equation}\label{4.37}
A=-\frac{2\,\cosh W}{1-\bar E^2}\,\bar E_{\bar u}\ ,
\qquad B=-\frac{2\,\cosh W}{1-\bar E^2}\,\bar E_{\bar v}\ .
\end{equation}
Substitution into (\ref{4.30}) simplifies this equation to 
\begin{equation}\label{4.38}
E_{\bar u}-E_{\bar v}=i\,k\,\lambda\,\left (E_{\bar u}+
E_{\bar v}\right )\ .\end{equation}
With $\lambda$ given by (\ref{3.35}) this equation establishes 
that 
\begin{equation}\label{4.39}
E=E(w)\qquad {\rm with}\qquad w=\sin (\bar u+\bar v)+i\,k\,
\sin (\bar u-\bar v)\ .\end{equation}
We can now determine $E$ using the boundary conditions 
(\ref{4.35}) and (\ref{4.36}). To see this easily we first write 
$k$ in (\ref{4.15}) in the form
\begin{equation}\label{4.40}
k=-i\,\frac{\left ({\rm e}^{-i\hat\alpha}-{\rm e}^{-i\hat\beta}
\right )}{{\rm e}^{-i\hat\alpha}+{\rm e}^{-i\hat\beta}}\ .\end{equation}
Using this in (\ref{4.39}) we see that we can consider $E$ to have 
the functional dependence:
\begin{equation}\label{4.41}
E=E\left ({\rm e}^{-i\hat\alpha}\sin\bar u\,\cos\bar v+
{\rm e}^{-i\hat\beta}\cos\bar u\,\sin\bar v\right )\ .\end{equation}
Now the boundary conditions (\ref{4.35}) and (\ref{4.36}) 
establish that
\begin{equation}\label{4.42}
E={\rm e}^{-i\hat\alpha}\sin\bar u\,\cos\bar v+
{\rm e}^{-i\hat\beta}\cos\bar u\,\sin\bar v\ ,\end{equation}
and thus the functions $V, W$ appearing in the line--element 
(\ref{2.2}) are determined by (\ref{4.33}) in the 
coordinates $(\bar u, \bar v)$. They are then converted into 
the coordinates $(u, v)$ using the transformations (\ref{3.30}) 
(see section 5 below).

Finally in the barred system the field equations (\ref{2.14}) 
and (\ref{2.15}) for $\bar M$ read (using (\ref{4.33}) 
and (\ref{4.37}))
\begin{eqnarray}\label{4.43}
\bar M_{\bar u}&=& -\frac{U_{\bar u\bar u}}{U_{\bar u}}
+\frac{1}{2}\,U_{\bar u}+\frac{2\,E_{\bar u}
\bar E_{\bar u}}{U_{\bar u}\,\left 
(1-\left |E\right |^2\right )^2}\ ,\\
\label{4.44}
\bar M_{\bar v}&=& -\frac{U_{\bar v\bar v}}{U_{\bar v}}
+\frac{1}{2}\,U_{\bar v}+\frac{2\,E_{\bar v}\bar E_
{\bar v}}{U_{\bar v}\,\left 
(1-\left |E\right |^2\right )^2}\ ,\end{eqnarray}
with $\bar M$ related to $M$ by (\ref{3.32}). Since 
we must have $M=0$ when $u=0$ and when $v=0$ we 
see from (\ref{3.32}) that
\begin{equation}\label{4.45}
{\rm when}\ \ \bar u=0\ ,\qquad {\rm e}^{-\bar M}=
\frac{\cos\bar v}{\sqrt{\alpha ^2+\beta ^2}\,
\sqrt{a^2+b^2}}\ ,\end{equation}
and 
\begin{equation}\label{4.46}
{\rm when}\ \ \bar v=0\ ,\qquad {\rm e}^{-\bar M}=
\frac{\cos\bar u}{\sqrt{\alpha ^2+\beta ^2}\,
\sqrt{a^2+b^2}}\ .\end{equation}
In (\ref{4.43}) and (\ref{4.44}), $U$ is given by (\ref{3.33})
and $E$ by (\ref{4.42}). Using (\ref{4.42}) we 
find
\begin{equation}\label{4.47}
E_{\bar u}\,\bar E_{\bar u}=1-\left |E\right |^2=E_{\bar v}\,\bar E_{\bar v}\ ,
\end{equation}
with 
\begin{equation}\label{4.48}
1-\left |E\right |^2=\cos ^2\left (\frac{\hat\alpha 
-\hat\beta}{2}\right )\,\cos ^2(\bar u+\bar v)+
\sin ^2\left (\frac{\hat\alpha 
-\hat\beta}{2}\right )\,\cos ^2(\bar u-\bar v) .
\end{equation}
The only complication in solving (\ref{4.43}) and (\ref{4.44}) 
is in dealing with the final term in each. In the case of (\ref{4.43}) 
this now involves evaluating the integral
\begin{equation}\label{4.49}
\int\frac{2\,d\bar u}{U_{\bar u}\left (1-\left |E\right |^2\right )}=
\int\frac{2\,\lambda\,d\lambda}{(\lambda ^2-1)\,\left\{
\cos ^2\left (\frac{\hat\alpha-\hat\beta}{2}\right )+
\lambda ^2\,\sin ^2\left (\frac{\hat\alpha-\hat\beta}{2}\right )
\right\}}\ ,\end{equation}
where we have changed the variable of integration from $\bar u$ 
to $\lambda$, given in (\ref{3.35}), with $\bar v$ held fixed. 
This integral is easy to evaluate and using (\ref{4.48}) again 
we obtain from (\ref{4.43})
\begin{equation}\label{4.50}
{\rm e}^{-\bar M}=\frac{1-\left |E\right |^2}{F(\bar v)\,\sqrt{
\cos (\bar u-\bar v)\,\cos (\bar u +\bar v)}}\ ,\end{equation}
with $F(\bar v)$ a function of integration. By (\ref{4.45}) we 
find that in fact $F$ is a constant given by
\begin{equation}\label{4.51}
F=\sqrt{\alpha ^2+\beta ^2}\,\sqrt{a^2+b^2}\ .\end{equation}
It is straightforward to see that (\ref{4.50}) with 
(\ref{4.51}) also satisfies (\ref{4.44}). The integration 
of Einstein's vacuum field equations is now complete.

\setcounter{equation}{0}
\section{Discussion}\indent
The purpose of this paper has been to propose a 
simple key to open up the boundary--value problem 
which is involved in deriving a model in General 
Relativity of the vacuum gravitational field left 
behind after the head--on collision of two plane 
impulsive gravitational waves. This `key' has 
been provided, with some motivation, in section 3 
(following equation(\ref{3.23})) and a physical 
interpretation has been suggested there for the 
interesting equation (\ref{3.43}), which is a 
consequence of the key assumption and some of the 
vacuum field equations. Our approach has been to 
focus attention on properties of the backscattered 
gravitational radiation present after the collision. 
This appears to be a non--linear phenomenon whose 
presence therefore ought to be expected to play a 
central role in the development of a scattering theory 
for gravitational radiation.

Notwithstanding the simplicity of our key assumption, 
the derivation of the line--element of the vacuum space--time 
in the region $\cal S$ in section 4 [the region $\cal S$ is 
defined following equation (\ref{4.28})] is complicated. 
It is therefore useful to summarise the result: the 
vacuum space--time in the region $\cal S$ has line--
element of the form (\ref{2.2}) with $U$ given by 
(\ref{2.17}), $V$ and $W$ given by (\ref{4.33}) 
with the complex function $E$ in (\ref{4.42}) expressed 
in coordinates $(u, v)$ as
\begin{equation}\label{5.1}
E=(\alpha +i\beta )\,u\,\sqrt{1-(a^2+b^2)\,v^2}+
(a+ib)\,v\,\sqrt{1-(\alpha ^2+\beta ^2)\,u^2}\ ,
\end{equation}
while $M$ is reconstructed in coordinates $(u, v)$ 
using (\ref{3.30}), (\ref{3.32}), (\ref{4.50}) and 
(\ref{4.51}). The result is
\begin{equation}\label{5.2}
{\rm e}^{-M}=\frac{\left (1-\left |E\right |^2\right )\,{\rm e}
^{U/2}}{\sqrt{1-(\alpha ^2+\beta ^2)\,u^2}\,\sqrt{
1-(a^2+b^2)\,v^2}}\ ,\end{equation}
with ${\rm e}^U$ given in (\ref{2.17}). If in (\ref{2.17}), 
(\ref{5.1}) and (\ref{5.2}) we put $a^2+b^2=1=\alpha ^2
+\beta ^2$ we recover the original form of the Nutku and 
Halil \cite{HN} solution. If in addition $b=\beta =0$ 
(and thus $E=\bar E$ and so $W=0$) we recover the 
original form of the Khan and Penrose \cite{KP} 
solution. We note that in the region $\cal S$ a 
curvature singularity \cite{KP}, \cite{HN} is 
encountered on the boundary where $(a^2+b^2)\,v^2+(\alpha ^2+\beta ^2)\,u^2=1$ 
and the solution above is valid only up to 
this space--like subspace.

Finally we wish to emphasise that the approach to solving 
collision problems in General Relativity developed in this 
paper is not confined to the examples worked through above. 
With different boundary conditions to those described 
by equations (\ref{2.3})--(\ref{2.10}) and the paragraph 
following (\ref{2.10}), the technique is capable 
of solving new collision problems (see, for example \cite{LMP}).  

This collaboration was funded by the Department of Education 
and Science and by the Minist\`ere des Affaires 
Etrang\`eres.

\appendix
\section{Plane Impulsive Wave}
\setcounter{equation}{0}
\noindent

The line--element of the vacuum space--time describing the 
gravitational field of an impulsive pp--wave \cite{EK} is 
given by 
\begin{equation}\label{A-1}
ds^2=-2dZ\,d\bar Z+2\,dV\,\left (dU+\delta\left (V\right )\,
f\left (Z,\bar Z\right )\,dV\right )\ .
\end{equation}
Here $f$ is a real--valued function which is harmonic in $Z, 
\bar Z$:
\begin{equation}\label{A-2}
\frac{\partial ^2f}{\partial Z\partial\bar Z}=0\ .
\end{equation}
The only non--vanishing component of the Riemann tensor in 
Newman--Penrose notation is 
\begin{equation}\label{A-3}
\Psi _0=\frac{\partial ^2f}{\partial Z^2}\,\delta\left (
V\right )\ ,\end{equation}
which is therefore Petrov type N with $\partial /\partial U$ as 
degenerate principal null direction. Integral curves of $\partial 
/\partial U$, which have vanishing expansion and shear, generate 
the history of the wave--front $V=0$. Hence $V=0$ is a null 
hyper{\it plane}. Following the example of Penrose \cite {P} 
a discontinuous coordinate transformation removes the $\delta$--
function from the line--element (\ref{A-1}) and introduces a 
coordinate system $(v, u, \zeta , \bar\zeta )$ in which the 
metric tensor is continuous $\left (C^0\right )$ across $V=0$. 
This transformation is given by
\begin{equation}\label{A-4}
V=v,\qquad U=u-\theta (v)\,f(\zeta , \bar\zeta )+|g|^2
v_+\ ,\qquad Z=\zeta +v_+\bar g(\bar \zeta )\ ,
\end{equation}
where $\theta (v)$ is the Heaviside step function, $v_+=v\,\theta (v)$ and $g(\zeta )
=-\partial f/\partial\zeta$ is an analytic function of $\zeta$. The 
line--element (\ref{A-1}) is transformed under (\ref{A-4}) into 
the Rosen form
\begin{equation}\label{A-5}
ds^2=-2\,\left |d\zeta +v_+\bar K(\bar\zeta )\,d\bar\zeta\right |^2+2\,du\,dv\ ,
\end{equation}
with $K(\zeta )=dg/d\zeta =-\partial ^2f/\partial\zeta ^2$ 
and (\ref{A-3}) becomes
\begin{equation}\label{A-6}
\Psi _0=\frac{\partial ^2f}{\partial\zeta ^2}\,\delta (v)\ .
\end{equation}
For a {\it plane} impulsive wave with two degrees of freedom of 
polarisation $\partial ^2f/\partial\zeta ^2$ is a complex 
constant. We take $f=-{\rm Re}\left\{(a+ib)\,\zeta ^2\right \}$ 
with $a, b$ real constants for this case. For a linearly polarised 
plane impulsive gravitational wave either $a=0$ or $b=0$. We note 
that (\ref{A-4}) incorporates Penrose's geometrical construction 
of the impulsive pp--wave whereby the history of the wave is 
formed in Minkowskian space--time by first subdividing the 
space--time into two halves $v>0$ and $v<0$ each with boundary 
$v=0$ and then reattaching the halves on $v=0$ identifying the 
points $(v=0, u, \zeta )$ and $(v=0, u-f(\zeta , \bar\zeta ), \zeta )$. 
This is a mapping of the two copies of $v=0$ where points on the 
same generators $\zeta ={\rm constant}$ of $v=0$ are mapped one 
to the other by the translation $u\rightarrow u-f(\zeta , \bar\zeta)$ 
and the mapping preserves the intrinsic degenerate metric on 
$v=0$.

\section{Incoming Waves Linearly Polarised}
\setcounter{equation}{0}
\noindent

If the incoming waves are linearly polarised $b=0$ 
in (\ref{2.3})--(\ref{2.5}) and $\beta =0$ in 
(\ref{2.7})--(\ref{2.9}). In the interaction region 
($u>0, v>0$) after the collision the function $W=0$. 
This is the problem solved by Khan and Penrose \cite{KP}. 
Our basic assumption following (\ref{3.23}) in 
this case reads: {\it there exist parameters ($\bar u, \bar v$) 
along the integral curves of $n$ and $l$ respectively 
such that $V_{\bar u}=V_{\bar v}$}. These parameters 
are given by (\ref{3.30}) with $b=\beta =0$. Now $V=V(\bar u+\bar v)$ 
in $u>0, v>0$. In the barred coordinates the boundary 
value of $V$ when $\bar u=0$ is obtained from (\ref{2.4}) 
with $b=0$ to be:
\begin{equation}\label{B-1}
{\rm e}^V=\frac{1+\sin\bar v}{1-\sin\bar v}\ .
\end{equation}
Thus for $u>0, v>0$ we have
\begin{equation}\label{B-2}
{\rm e}^V=\frac{1+\sin (\bar u+\bar v)}{1-\sin (\bar u+\bar v)}
=\frac{(\cos\bar u+\sin\bar v)(\cos\bar v+\sin\bar u)}{(\cos\bar u
-\sin\bar v)(\cos\bar v-\sin\bar u)}\ .
\end{equation}
Using (\ref{3.30}) with $b=\beta =0$ we can write (\ref{B-2}) 
as
\begin{equation}\label{B-3}
{\rm e}^V=\frac{(\sqrt{1-\alpha ^2u^2}+a\,v)(\sqrt{1-a^2v^2}+\alpha\,
u)}{(\sqrt{1-\alpha ^2u^2}-a\,v)(\sqrt{1-a^2v^2}-\alpha\,
u)}\ ,\end{equation}
which is the Khan and Penrose \cite{KP} expression for $V$ 
in the interaction region. From (\ref{3.30}) and (\ref{3.33}) 
we have $U$ in the coordinates $(u, v)$ given by
\begin{equation}\label{B-4}
{\rm e}^{-U}=1-\alpha ^2u^2-a^2v^2\ .
\end{equation}

The only remaining function to be determined is $M$. In 
the coordinates $(\bar u, \bar v)$ this is replaced by 
$\bar M$ with the latter related to $M$ via (\ref{3.32}) 
with $b=\beta =0$. If we let $\bar Q=\bar M+\frac{1}{2}U$ 
then in the barred coordinates the field equations 
(\ref{2.14}) and (\ref{2.15}) reduce to the following 
equations for $\bar Q$:
\begin{equation}\label{B-5}
\bar Q_{\bar u}=\bar Q_{\bar v}=2\,\tan (\bar u+\bar v)\ .
\end{equation}
Since $M=0$ when $u=0$ or $v=0$ it is easy to see that 
when $\bar u=0$, $\bar Q=\log\left (\cos ^2\bar v/\alpha\,a
\right )$ and when $\bar v=0$, $\bar Q=\log\left (\cos ^2\bar u/\alpha\,a
\right )$. It thus follows from (\ref{B-5}) that 
\begin{equation}\label{B-6}
\bar Q=-\log\left (\frac{\cos ^2(\bar u+\bar v)}{\alpha\,a}\right )\ .
\end{equation}
Hence by (\ref{3.32}) with $b=\beta =0$:
\begin{equation}\label{B-7}
{\rm e}^{-M}=\frac{\alpha\,a}{\cos\bar u\,\cos\bar v}\,
{\rm e}^{-\bar Q+\frac{1}{2}U}=\frac{[\cos (\bar u+\bar v)\,
\cos (\bar u-\bar v)]^{\frac{3}{2}}}{\cos\bar u\,\cos\bar v\,cos ^2(\bar u-\bar v)}\ .
\end{equation}
In the $(u, v)$ coordinates this reads
\begin{equation}\label{B-8}
{\rm e}^{-M}=\frac{(1-\alpha ^2u^2-a^2v^2)^{\frac{3}{2}}}
{\sqrt{1-\alpha ^2u^2}\,\sqrt{1-a^2v^2}[\sqrt{1-\alpha ^2u^2}\,
\sqrt{1-a^2v^2}+\alpha\,a\,u\,v]^2}\ ,
\end{equation}
which is the Khan and Penrose \cite{KP} expression.


\begin{thebibliography}{99}
\bibitem{EX} It is well known (see \cite{JG}) that (\ref{2.11}) results in ${\rm e}^{-U}=
f_1(u)+f_2(v)$ for some functions $f_1, f_2$ and these functions 
are immediately determined by the boundary values (\ref{2.3}) and 
(\ref{2.7}) of $U$.
\bibitem{INT} This follows because (\ref{4.23}) can 
be written $\frac{d}{d\lambda}\left [\left (\frac{1}{\lambda}
-\lambda\right )\,X\right ]=0$ where $X=\lambda\,P'+
\left (\frac{1+\lambda ^2}{1-\lambda ^2}\right )\,P-\frac{\lambda \,
G^2}{2(1-\lambda ^2)}$. The constant of integration 
then vanishes on account of (\ref{4.22}), which in 
particular implies by (\ref{4.7}) that $P(1)=0$.
\end{thebibliography}
\end
{document}